\documentclass[apjl]{emulateapj}
\usepackage{apjfonts}
\begin{document}
\title{Redshift Evolution of the Galaxy Velocity Dispersion Function}

\author{Rachel Bezanson\altaffilmark{1}, Pieter G. van Dokkum\altaffilmark{1}, Marijn Franx\altaffilmark{2}, Gabriel B. Brammer\altaffilmark{3}, Jarle Brinchmann\altaffilmark{2}, Mariska Kriek\altaffilmark{4}, Ivo Labb\'{e} \altaffilmark{2}, Ryan F. Quadri\altaffilmark{5,6}, Hans-Walter Rix\altaffilmark{7}, Jesse van de Sande\altaffilmark{2}, Katherine E. Whitaker\altaffilmark{1}, Rik J. Williams\altaffilmark{5}}\altaffiltext{1}{Department of Astronomy, Yale University, New Haven, CT 06520-8101}
\altaffiltext{2}{Sterrewacht Leiden, Leiden University, NL-2300 RA Leiden, Netherlands}
\altaffiltext{3}{European Southern Observatory, Alonso de C—rdova 3107, Casilla 19001, Vitacura, Santiago, Chile}
\altaffiltext{4}{Harvard-Smithsonian Center for Astrophysics, Cambridge, MA}
\altaffiltext{5}{Carnegie Observatories, Pasadena, CA 91101}
\altaffiltext{6}{Hubble Fellow}
\altaffiltext{7}{Max-Planck-Institut f\"{u}r Astronomie, K\"{o}nigstuhl 17, D-69117 Heidelberg, Germany}

\slugcomment{Accepted for publication in ApJ Letters}
\shortauthors{Bezanson et al.}
\shorttitle{The Evolving Galaxy VDF}

\newcommand{\unit}[1]{\ensuremath{\, \mathrm{#1}}}

\begin{abstract}

We present a study of the evolution of the galaxy Velocity Dispersion
Function (VDF) from $z=0$ to $z=1.5$ using photometric data from the UDS and NMBS COSMOS
surveys.  The VDF has been measured locally using direct kinematic
measurements from the Sloan Digital Sky Survey, but direct studies of
the VDF at high redshift are difficult as they require velocity
dispersion measurements of many thousands of galaxies.
{Taylor} {et~al.} (2010) demonstrated that dynamical and stellar mass are
linearly related when the structure of the galaxy is accounted for.
We show that the stellar mass, size and S\'ersic index can reliably
predict the velocity dispersions of SDSS galaxies.  We apply this
relation to galaxies at high redshift and determine the evolution
of the inferred VDF. We find that the VDF at $z\sim0.5$ is very similar to the
VDF at $z=0$.  At higher redshifts, we find that the
number density of galaxies with dispersions $\lesssim200\unit{km/s}$
is lower, but the number of high dispersion galaxies is
constant or even higher.  At fixed cumulative
number density, the velocity dispersions of galaxies with
$\log\,N[\unit{Mpc^{-3}}]<-3.5$ increase with time by a factor of
$\sim1.4$ from $z\sim1.5-0$, whereas the dispersions of galaxies with
lower number density are approximately constant or decrease with time.
The VDF appears to show less evolution than the stellar
mass function, particularly at the lowest number densities.
We note that these results
are still somewhat uncertain and
we suggest several avenues for further calibrating the
inferred velocity dispersions.
\end{abstract}

\keywords{cosmology: observations --- galaxies: evolution --- galaxies:
formation --- galaxies: elliptical and lenticular, cD}

\section{Introduction}

Stellar velocity dispersion is a fundamental property of galaxies.
Through $M\propto\,R_{\star}\sigma_{\star}^2/G$ it provides a
characterization of the galaxy mass and is a key axis in the
Fundamental Plane of early-type galaxies
(e.g., {Djorgovski} \& {Davis} 1987; {Bernardi} {et~al.} 2003).  Further,
it appears to correlate strongly with many other properties of galaxies,
such as specific star formation rates, galaxy color, and
black hole mass ({Magorrian} {et~al.} 1998; {Franx} {et~al.} 2008; {Trujillo}, {Ferreras}, \& {de la  Rosa} 2011).

The evolution of an individual galaxy's velocity dispersion carries
information about the physical mechanisms responsible for its
growth. Broadly, processes that increase mass more efficiently than
size, such as central gas accretion and the resulting star formation,
could increase velocity while processes such as minor merging or mass
loss in galactic winds could increase the overall size of a galaxy and
decrease the velocity dispersion. The central dispersions may also
be fairly stable with time and
reflect the central dark matter
potentials at the time when the galaxy started
forming ({Loeb} \& {Peebles} 2003).  In this scenario, high
redshift galaxies form the central regions of local massive galaxies
and still retain signatures of their earliest progenitors.

Furthermore, central velocity dispersion lies at the intersection of
observational properties and quantities predictable from simulations.
Observational evidence about the shape and evolution of the VDF can be
directly compared to predictions from cosmological simulations.
Finally, masses of the supermassive black holes in the centers of
galaxies display the strongest correlations with host galaxy velocity
dispersions -- understanding the distribution of velocity dispersions
would hold further significance for the study of black holes and AGN.

In the local universe, the VDF has been measured directly (e.g., {Sheth} {et~al.} 2003; {Mitchell} {et~al.} 2005; {Choi}, {Park}, \& {Vogeley} 2007).  While measurements of stellar velocity dispersion are possible out to $z\sim1$, they become prohibitive for large samples at $z>1$.  At $z=0$ we can predict velocity dispersions of galaxies very well from their photometric properties.  Here we present a new approach to describe the VDF based on photometric predictions and constrain its evolution out to $z\sim1.5$.  For this work we assume a concordance cosmology ($H_0=70\rm{\,km/s\,Mpc^{-1}},\Omega_M=0.3\,\&\,\Omega_{\Lambda}=0.7$).

\section{Data}
	
\subsection{Spectroscopic Redshifts, Velocity Dispersions, Sizes and Stellar Masses at $z\sim0$}

The $z\sim0$ data is based on analysis of several publicly available
catalogs from SDSS DR7.  Galaxies are included from $0.05<z<0.07$,
using photometric information from the main DR7 catalogs ({Abazajian} {et~al.} 2009)
and redshift and velocity dispersions from the Princeton pipeline.
The sample is selected to have good photometric measurements with the
SDSS Science Primary flag, and low relative errors in velocity
dispersion ($<10\%$).  All velocity dispersions are aperture corrected
to $r_e/8$ using $\sigma_0=\sigma_{ap}(8.0r_{ap}/r_{e})^{0.066}$ based
on the best-fit correction to the SAURON sample ({Cappellari} {et~al.} 2006),
where $r_{ap}=1''.5$ is the radius of the SDSS spectroscopic fiber.
We adopt the best-fit {S\'ersic} (1968) effective radii in the $r'$ band
from the NYU-VAGC ({Blanton} {et~al.} 2005).  These sizes are based on fits
to azimuthally averaged light profiles and are equivalent to
circularized effective radii.  Stellar masses are computed by the
MPA-JHU group from the DR7 best-fit model
magnitudes\footnote{http://www.mpa-garching.mpg.de/SDSS/DR7/} ({Brinchmann} {et~al.} 2004).  We
adopt these $M_{\star}/L$ ratios and derive stellar masses from the
luminosity of the best-fit S\'ersic model for all galaxies.

\subsection{Sizes and Stellar Masses at $z>0$}

We use two samples of high redshift galaxies in this paper.  The first
is the $0.77\,\rm{deg^{2}}$ UKIDSS Ultra-Deep Survey (UDS) K-selected
galaxy catalog ({Williams} {et~al.} 2009, 2010).  This catalog includes
near-infrared photometry (JHK) from the UKIDSS UDS
({Lawrence} {et~al.} 2007; {Warren} {et~al.} 2007), optical imaging from the SXDS
({Sekiguchi} \& {SXDS} 2004) and $3.6/4.5\mu m$ data from the SWIRE survey
({Lonsdale} {et~al.} 2003).  Circularized effective radii ($r_e=\sqrt{ab}$)
were measured in the J, H and K images for all bright sources
($K<22.4$) from {S\'ersic} (1968) models fitted with
GALFIT ({Peng} {et~al.} 2002).
A full description of the catalog and size fitting as well as
associated tests can be found in {Williams} {et~al.} (2010).

\begin{figure}[t]
  \centering
%  \epsscale{0.5}
  \plotone{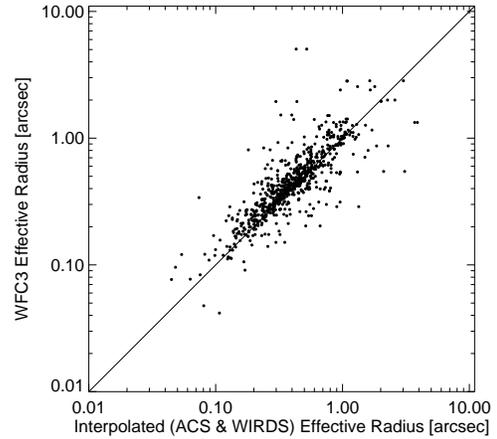}
  \caption{Sizes measured from WFC3 F140W imaging compared to sizes interpolated between measurements from ACS F814W and WIRDS Ks imaging.}
  \label{fig:groundsizetest}
\end{figure}

The second is the NEWFIRM Medium Band Survey (NMBS) of the COSMOS
field, which we briefly summarize below.  For an in-depth description
of the survey see {Whitaker} {et~al.} (2011).  The $0.21\unit{deg^2}$ field
includes medium-band NIR imaging ($J1,J2,J3,H1,H2,K$) from the Mayall
4m telescope, optical imaging ($ugriz$) from the CFHT Legacy Survey
and Subaru, IRAC imaging and MIPS $24\unit{\mu\,m}$ data.  Sizes for
the Cosmos galaxies are measured from the Ks WIRCam Deep Survey
(WIRDS) $0''.186/\unit{pixel}$ images (Bielby et al., in prep), which
have seeing of $\sim0".6-0".7$ and the v.1.3 ACS F814W mosaic from
{Scoville} {et~al.} (2007).  For details of galaxy modeling see van Dokkum,
et al., in prep. - we summarize as follows.  Circularized effective
radii, axis ratio and S\'ersic indices are determined for all galaxies
brighter than $K<22$ using S\'ersic models convolved with a
position-dependent PSF with GALFIT ({Peng} {et~al.} 2002).  We test our
procedure using independent high spatial resolution data from a small
area of overlapping WFC3 imaging from the 3D-HST survey (van Dokkum et
al., in prep).  The interpolated WIRDS/ACS sizes are consistent with
the WFC3 sizes down to $\sim0''.15$ with a biweight mean of
$\log\left(r_{e,\rm{ACS\&WIRDS}}/r_{e,\rm{WFC3}}\right)=0''.0009$
and a scatter of 0.1 dex
(Figure \ref{fig:groundsizetest}).

In both fields sizes are measured down to an arbitrary limit; however,
as most of these sizes are based on ground-based imaging we
assign a minimum size for all galaxies of $r_e=0''.2$.

Photometric redshifts for all galaxies in the UDS and COSMOS surveys
were calculated using the EAZY code ({Brammer}, {van Dokkum}, \& {Coppi} 2008).  Stellar masses were
computed using FAST ({Kriek} {et~al.} 2009), from {Bruzual} \& {Charlot} (2003) stellar
population synthesis models with Solar metallicity and a
{Chabrier} (2003) Initial Mass Function (IMF).

\section{Inferred Velocity Dispersion}

\begin{figure*}[!t]
  \centering
    \epsscale{1.1}
  \plottwo{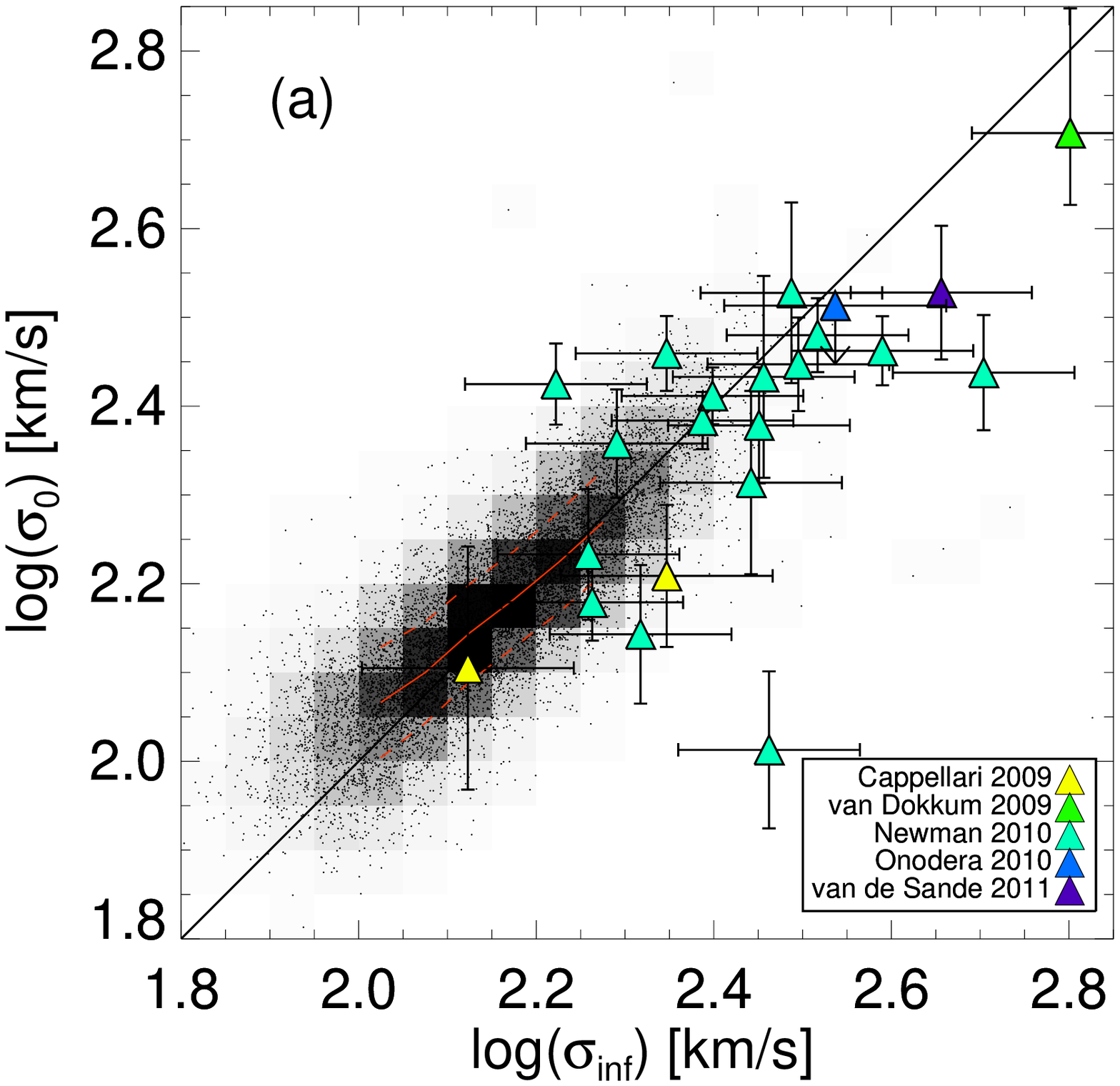}{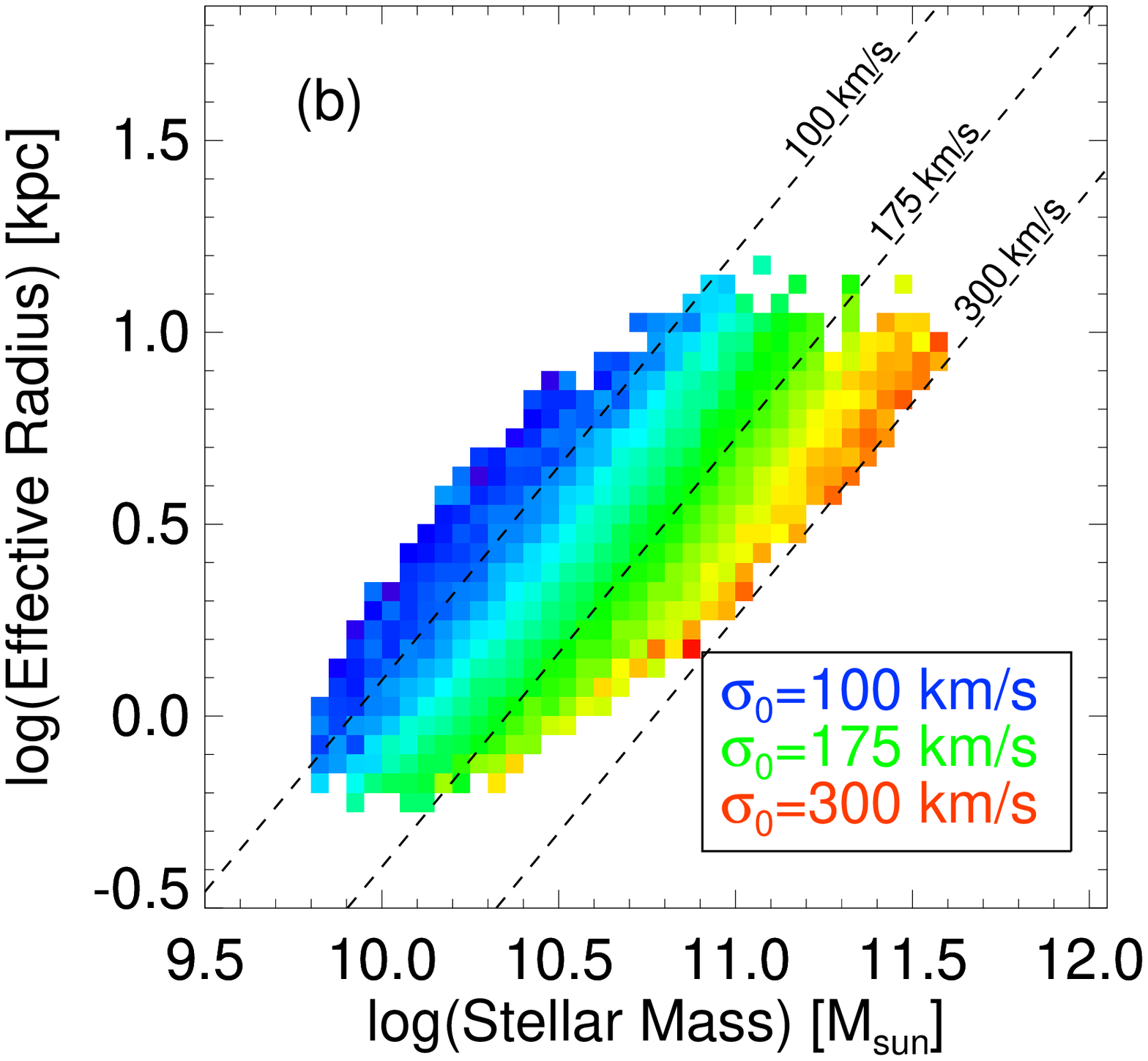}
  \caption{(a) Measured central velocity dispersion vs. inferred velocity dispersion of galaxies in the SDSS DR7.  Galaxies are binned by $\log{\sigma_{inf}}=0.05$ and the running mean and rms are shown in red for bins containing at least 100 galaxies.  There is a very good correlation with a scatter of $\sim0.06\unit{dex}$.  High redshift ($1<z<2.5$) galaxies are included as large colored triangles.  (b) Effective radii from NYU-VAGC best-fit S\'ersic models in the $r'$ band vs. Stellar Mass in the SDSS, color coded by median measured velocity dispersion.  Lines of constant inferred dispersion are included as dashed black lines.  Both panels illustrate that the stellar mass and size of galaxies can be used to predict velocity dispersions.}
  \label{fig:SDSS}
\end{figure*}

To derive a VDF at $z>0$, we now describe how to estimate the stellar velocity dispersion of galaxies from their photometric properties.  According to the virial theorem, the dynamical mass of a galaxy is proportional to the product of the square of its measured dispersion and effective radius.

\begin{equation}
M_{dyn}=K_v\sigma_0^2r_e/G
\end{equation}

Stellar mass has been shown to be proportional to dynamical mass for local galaxies (e.g., {Taylor} {et~al.} 2010).  However, {Taylor} {et~al.} (2010) demonstrated that simple estimates of dynamical mass based on homology exhibit residual trends with galaxy structural properties and introduced a more robust structure-corrected dynamical mass, which we use to calculate inferred velocity dispersion based on photometric estimates of stellar mass and observed size.  We adopt a S\'ersic dependent virial constant, $K_v(n)$ ({Bertin}, {Ciotti}, \& {Del  Principe} 2002)

\begin{equation}
K_v(n)=\frac{73.32}{10.465+(n-0.94)^2}+0.954.
\end{equation}

In order to predict the central velocity dispersion of a galaxy based on its size and stellar mass, we define 
\begin{equation}
\sigma_{inf}=\sqrt{\frac{GM_{dyn}}{K_v(n) r_e}}=\sqrt{\frac{GM_{\star}}{K_{\star}(n)r_e}},
\end{equation}
where
$K_{\star}(n)\equiv\,K_v(n)\left(\frac{M_{\star}}{M_{dyn}}\right)$.
Since {Taylor} {et~al.} (2010) showed that
$K_v(n)\left(\frac{M_{\star}}{M_{dyn}}\right)$ depends only weakly on
mass, we adopt the average ratio of stellar to total mass,
calibrated such that the median $\sigma_{inf}=\sigma_{0}$ for
$2.0<\log\sigma_0<2.4$ in the SDSS.  We find that
$\left<\frac{M_{\star}}{M_{dyn}}\right>=0.557$, therefore
$K_{\star}(n)=0.557K_v(n)$.  The central and inferred velocity
dispersions agree well for the SDSS, with a $1-\sigma$ error of
$0.06\unit{dex}$ (Figure \ref{fig:SDSS}a).  This is remarkable,
particularly when considering that no morphological selections are
applied: after applying the S\'ersic dependent virial constant, this relation holds for both ellipticals and spiral galaxies.
This implies that the velocity dispersion, $\sigma_0$, can be well
predicted from $M_{\star}$ and $r_e$.  Figure \ref{fig:SDSS}b shows
the size-mass relation for SDSS, colored by median measured dispersion
for bins including a minimum of five galaxies.  We use the median
$n(M_{\star})$ to overplot lines of constant inferred dispersion.
Again, the measured and inferred dispersions agree quite well.

Also included in Figure \ref{fig:SDSS}a are published measurements for
galaxies  at $1.0<z<2.5$
({Cappellari} {et~al.} 2009; {van Dokkum}, {Kriek}, \&  {Franx} 2009; {Newman} {et~al.} 2010; {Onodera} {et~al.} 2010; {van de Sande} {et~al.} 2011)
(large colored triangles), assuming errors due to SED modeling of
$M_{\star}=0.2\unit{dex}$.  The biweight mean offset in $\sigma_0/\sigma_{inf}$ is small ($-0.056\pm0.025\unit{dex}$) and the high observed scatter of $0.131\pm0.028\unit{dex}$ is fully consistent with scatter due to measurement errors.  The relation does not appear to evolve
significantly out to high redshift, but we emphasize that
current samples of high-$z$ dynamical
measurements are small and biased toward high dispersions.  Motivated by Figure
\ref{fig:SDSS}a we assume that the SDSS relations
between velocity dispersion, stellar mass, effective radius
and S\'ersic index do not evolve with
redshift and calculate inferred dispersions for the UDS and COSMOS
catalogs.  To limit bandpass-dependent effects we calculate
$\sigma_{inf}$ separately using UKIDSS J and K sizes for the UDS and
ACS F814W and WIRDs Ks sizes for Cosmos.  Next, we interpolate
$\sigma_{inf}$ to a rest-frame wavelength of $6200\unit{\AA}$, the
central wavelength of the $r'$ filter.  We note that for apparently
extremely compact galaxies with measured sizes $<0''.2$, these can be
interpreted as minimum inferred velocity dispersions.

\section{Velocity Dispersion Function}

\begin{figure*}[!t]
  \centering
  \epsscale{2.5}
  \plottwo{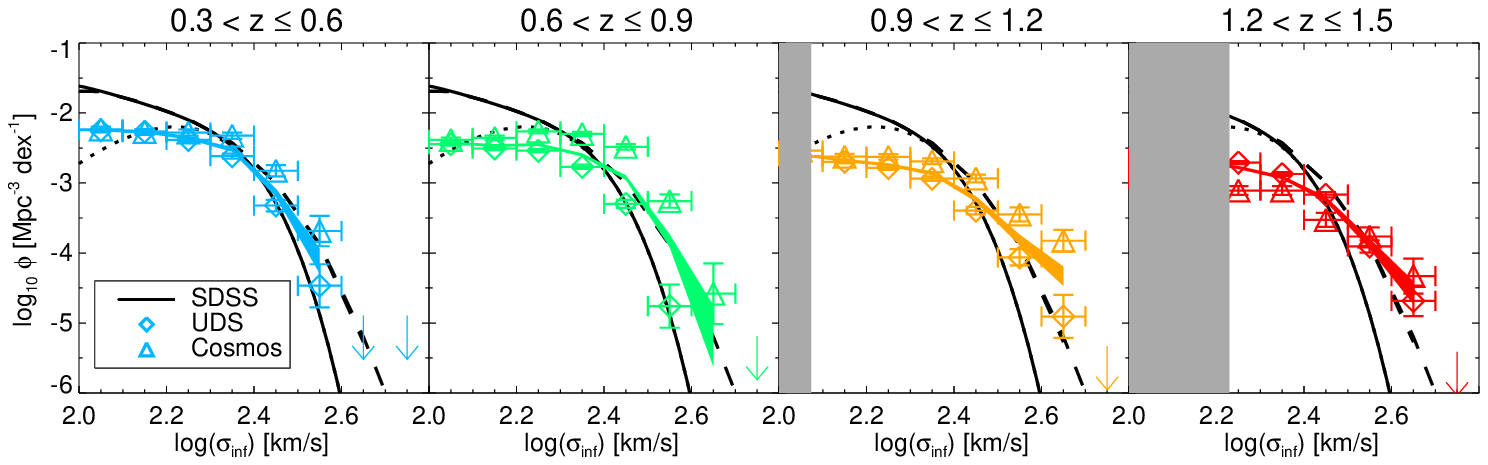}{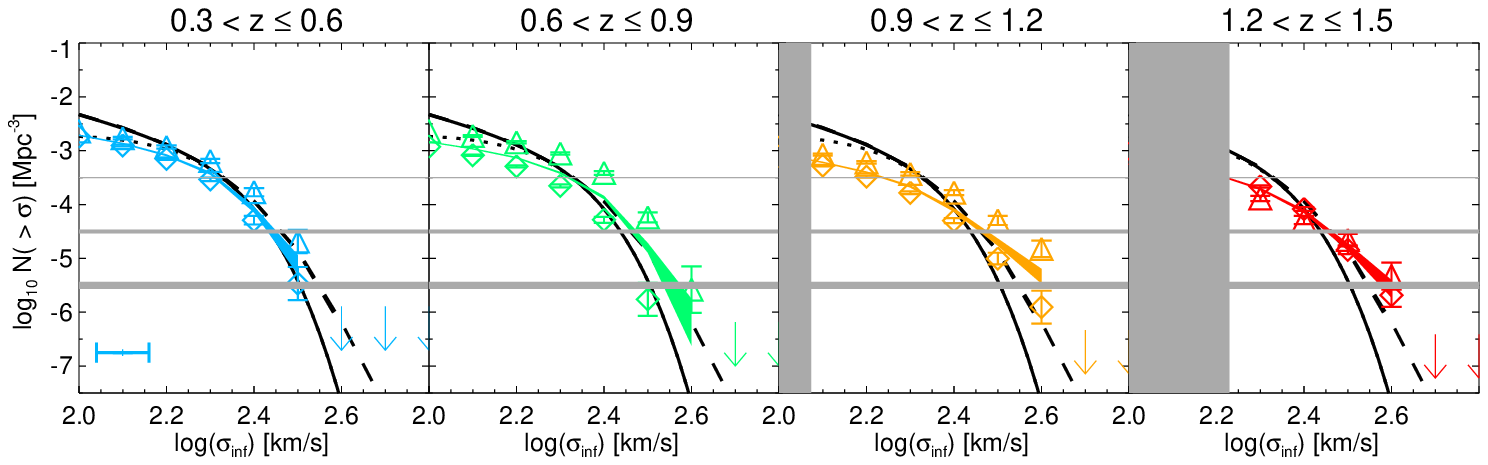}
  \caption{\emph{Top row}: VDFs locally (black) as measured from SDSS and at higher redshifts for the two fields (color).  Best-fit modified {Schechter} (1976) fits to the SDSS measured VDF are shown from {Sheth} {et~al.} (2003) (\emph{thin dotted} - early type VDF, \emph{thick solid} - total VDF, \emph{thick dashed} - total scattered VDF).  Higher redshift number densities are based on inferred velocity dispersions assuming $z=0$ calibration.  Individual colored lines reflect the measured VDF in each field (solid for UDS and dashed for Cosmos) and the circular data points show the volume-weighted average.  \emph{Bottom row}: Cumulative VDFs for each redshift bin.  Colors and symbols are the same as above.  Three cuts in number density (\emph{horizontal gray lines}) highlight three possible evolutionary behaviors.  Galaxies at the lowest number density threshold, $\log\,N[\unit{Mpc^{-3}}]<-5.5$, increase in velocity dispersion from a redshift $z\sim1.5$ to $z\sim0$.  Galaxies at a number density of $\log\,N[\unit{Mpc^{-3}}]<-4.5$ appear to be near the crossover and experience little evolution in velocity dispersion to $z\sim0$.  Finally galaxies at the highest number density threshold of $\log\,N[\unit{Mpc^{-3}}]<-3.5$ appear to decrease in velocity dispersion with time.}
  \label{fig:VDFcomp}
\end{figure*}

A number of estimates of the $z\sim0$ VDF of early-type galaxies exist
based on the SDSS ({Sheth} {et~al.} 2003; {Mitchell} {et~al.} 2005; {Choi} {et~al.} 2007; {Chae} 2010).  These
studies are in reasonably good agreement and show that the $z=0$ VDF
appears to be well fit by modified {Schechter} (1976) functions.
{Sheth} {et~al.} (2003) present a correction to the VDF for the contribution
due to late-type galaxies, based on circular velocities.  In the top
row of Figure \ref{fig:VDFcomp} we show both the fit to the early-type
VDF (thin dotted black lines) and the total VDF (thick solid black lines), which
includes the correction for late-type galaxies.  This latter function
is calculated by fitting a polynomial to the data in Figure 6 in
{Sheth} {et~al.} (2003).

Measurement errors of velocity dispersions have a non-negligible
effect on the shape of the inferred VDF,  particularly on the steep,
high dispersion tail.  We model this effect by producing a mock
catalog that intrinsically follows the total {Sheth} {et~al.} (2003) VDF.  In
$\S3$, we measured a scatter of $0.06\unit{dex}$ in inferred
dispersion, which is comprised of intrinsic scatter in this relation
combined with measurement errors in velocity dispersion.  From the
median relative error in velocity dispersion for galaxies
with $\sigma\ge200\unit{km/s}$ ($0.028\unit{dex}$) 
we estimate that the intrinsic scatter in the relation between
measured and inferred dispersion is
$0.053\unit{dex}$. The thick dashed black lines in
Fig.\ \ref{fig:VDFcomp} show the SDSS VDF when this
scatter is added to the mock velocity dispersions.

The top panel of Figure \ref{fig:VDFcomp} shows the inferred VDF for
each field in all redshift ranges and values are included in Table
\ref{tbl:VDF}.  Errors reflect Poisson errors in each velocity
dispersion bin.
Incompleteness is mostly caused by galaxies that are too faint to
measure reliable sizes. We define the dispersion completeness
limit (grey regions in Figure \ref{fig:VDFcomp}) for each field and
redshift range as the dispersion at which the $95\%$ completeness
plus $0.06\unit{dex}$ scatter about a linear fit to the K magnitude -
inferred dispersion relation reaches the magnitude limit of the size
catalogs.

In the lowest redshift range, $0.3<z<0.6$, the observed VDF matches
the local VDF quite well in the region dominated by early-type
galaxies, especially when compared to the scattered VDF.  At low
dispersions the $z\sim0.5$ function is intermediate between the $z=0$
VDF for early-types only (black dashed line) and for all galaxies
(black solid line).  In the higher redshift bins, the VDF appears to
flatten.  Low dispersion galaxies become less common and high
dispersion galaxies more common, with an approximately constant
crossover point.  The colored points show results for the UDS
(diamonds) and COSMOS (triangles) fields separately.  The flattening
of the VDF with redshift is seen in both fields, suggesting that it is
not due to field-to-field variations.  We note that the COSMOS field
has a cluster at z=0.7 and a overdensity at $z\sim1.0$
({Whitaker} {et~al.} 2011).  These structures are reflected in the higher
normalization of the COSMOS VDF in the second and third redshift bins.

\begin{deluxetable*}{cccccccccc}
\tabletypesize{\scriptsize}
\tablecaption{Inferred  and Cumulative Velocity Dispersion Functions\label{tbl:VDF}}
\tablewidth{0pt}
\tablehead{
\colhead{$\log\sigma_{inf}$ [km/s]} & & \colhead{$2.0-2.1$} & \colhead{$2.1-2.2$} & \colhead{$2.2-2.3$} & \colhead{$2.3-2.4$}
 & \colhead{$2.4-2.5$} & \colhead{$2.5-2.6$} & \colhead{$2.6-2.7$} \\ [+0.5ex]
 \hline
 \hline \\[-1ex]
\colhead{Redshifts} & \colhead{Field} &  \multicolumn{7}{c}{$\log\Phi [\rm{Mpc}^{-3} \rm{dex}^{-1}]$}}
\startdata
\\ [-1ex]
$ 0.3 < z \leq 0.6$ & UDS &  $-2.25\pm 0.02$ &  $-2.27\pm 0.02$ &  $-2.38\pm 0.03$ &  $-2.62\pm 0.04$ &  $-3.32\pm 0.08$ &  $-4.47\pm 0.31$ &  $<-4.77$ \\ 
 & Cosmos &  $-2.24\pm 0.04$ &  $-2.27\pm 0.04$ &  $-2.28\pm 0.04$ &  $-2.32\pm 0.05$ &  $-2.83\pm 0.08$ &  $-3.69\pm 0.22$ &  $<-4.29$ \\ 
[+1ex]
 & Total &  $-2.24\pm 0.02$ &  $-2.27\pm 0.02$ &  $-2.36\pm 0.02$ &  $-2.52\pm 0.03$ &  $-3.14\pm 0.06$ &  $-4.11\pm 0.18$ &  $<-4.89$ \\ 
\\
$ 0.6 < z \leq 0.9$ & UDS &  $-2.44\pm 0.02$ &  $-2.51\pm 0.02$ &  $-2.54\pm 0.02$ &  $-2.77\pm 0.03$ &  $-3.30\pm 0.06$ &  $-4.76\pm 0.31$ &  $<-5.06$ \\ 
 & Cosmos &  $-2.38\pm 0.03$ &  $-2.36\pm 0.03$ &  $-2.26\pm 0.03$ &  $-2.30\pm 0.03$ &  $-2.48\pm 0.04$ &  $-3.26\pm 0.09$ &  $-4.58\pm 0.43$ \\ 
[+1ex]
 & Total &  $-2.43\pm 0.02$ &  $-2.46\pm 0.02$ &  $-2.45\pm 0.02$ &  $-2.60\pm 0.02$ &  $-2.92\pm 0.03$ &  $-3.82\pm 0.09$ &  $-5.19\pm 0.43$ \\ 
\\
$ 0.9 < z \leq 1.2$ & UDS &  $-2.62\pm 0.02$ &  $-2.70\pm 0.02$ &  $-2.78\pm 0.03$ &  $-2.94\pm 0.03$ &  $-3.40\pm 0.05$ &  $-4.06\pm 0.12$ &  $-4.91\pm 0.31$ \\ 
 & Cosmos &  $-2.54\pm 0.03$ &  $-2.62\pm 0.04$ &  $-2.63\pm 0.04$ &  $-2.69\pm 0.04$ &  $-2.93\pm 0.06$ &  $-3.45\pm 0.10$ &  $-3.82\pm 0.15$ \\ 
[+1ex]
 & Total &  $-2.60\pm 0.02$ &  $-2.68\pm 0.02$ &  $-2.74\pm 0.02$ &  $-2.86\pm 0.03$ &  $-3.23\pm 0.04$ &  $-3.81\pm 0.08$ &  $-4.33\pm 0.14$ \\ 
\\
$ 1.2 < z \leq 1.5$ & UDS &  $-2.65\pm 0.02$ &  $-2.64\pm 0.02$ &  $-2.71\pm 0.02$ &  $-2.87\pm 0.03$ &  $-3.17\pm 0.04$ &  $-3.91\pm 0.09$ &  $-4.68\pm 0.22$ \\ 
 & Cosmos &  $-2.94\pm 0.05$ &  $-2.92\pm 0.05$ &  $-3.11\pm 0.06$ &  $-3.11\pm 0.06$ &  $-3.53\pm 0.10$ &  $-3.76\pm 0.13$ &  $-4.33\pm 0.25$ \\ 
[+1ex]
 & Total &  $-2.70\pm 0.02$ &  $-2.69\pm 0.02$ &  $-2.78\pm 0.02$ &  $-2.92\pm 0.02$ &  $-3.23\pm 0.04$ &  $-3.87\pm 0.07$ &  $-4.57\pm 0.16$ \\ 
\\
\hline
\hline
\colhead{$\log\sigma_{inf}$ [km/s]} & & \colhead{$ >2.0$} & \colhead{$>2.1$} & \colhead{$>2.2$} & \colhead{$>2.3$}
 & \colhead{$>2.4$} & \colhead{$>2.5$} & \colhead{$>2.6$} \\ [+0.5ex]
 \hline
 \hline \\[-1ex]
\colhead{Redshifts} & \colhead{Field} &  \multicolumn{7}{c}{$\log\,N [\rm{Mpc}^{-3}]$}
\\[+0.5ex]
\hline
\\
$ 0.3 < z \leq 0.6$ & UDS &  $-1.74\pm 0.01$ &  $-1.90\pm 0.02$ &  $-2.15\pm 0.02$ &  $-2.53\pm 0.03$ &  $-3.29\pm 0.08$ &  $-4.47\pm 0.31$ &  $<-4.77$ \\ 
 & Cosmos &  $-1.64\pm 0.02$ &  $-1.77\pm 0.02$ &  $-1.93\pm 0.03$ &  $-2.19\pm 0.04$ &  $-2.77\pm 0.08$ &  $-3.69\pm 0.22$ &  $<-4.29$ \\ 
[+1ex]
 & Total &  $-1.71\pm 0.01$ &  $-1.87\pm 0.01$ &  $-2.09\pm 0.02$ &  $-2.42\pm 0.03$ &  $-3.09\pm 0.05$ &  $-4.11\pm 0.18$ &  $<-4.89$ \\ 
\\
$ 0.6 < z \leq 0.9$ & UDS &  $-1.93\pm 0.01$ &  $-2.08\pm 0.01$ &  $-2.29\pm 0.02$ &  $-2.65\pm 0.03$ &  $-3.28\pm 0.06$ &  $-4.76\pm 0.31$ &  $<-5.06$ \\ 
 & Cosmos &  $-1.64\pm 0.01$ &  $-1.73\pm 0.02$ &  $-1.84\pm 0.02$ &  $-2.05\pm 0.02$ &  $-2.41\pm 0.04$ &  $-3.24\pm 0.09$ &  $-4.58\pm 0.43$ \\ 
[+1ex]
 & Total &  $-1.84\pm 0.01$ &  $-1.96\pm 0.01$ &  $-2.13\pm 0.01$ &  $-2.41\pm 0.02$ &  $-2.87\pm 0.03$ &  $-3.81\pm 0.09$ &  $-5.19\pm 0.43$ \\ 
\\
$ 0.9 < z \leq 1.2$ & UDS &  $-2.11\pm 0.01$ &  $-2.28\pm 0.01$ &  $-2.48\pm 0.02$ &  $-2.78\pm 0.03$ &  $-3.30\pm 0.05$ &  $-4.00\pm 0.11$ &  $-4.91\pm 0.31$ \\ 
 & Cosmos &  $-1.94\pm 0.02$ &  $-2.07\pm 0.02$ &  $-2.22\pm 0.02$ &  $-2.43\pm 0.03$ &  $-2.78\pm 0.05$ &  $-3.30\pm 0.08$ &  $-3.82\pm 0.15$ \\ 
[+1ex]
 & Total &  $-2.06\pm 0.01$ &  $-2.22\pm 0.01$ &  $-2.40\pm 0.01$ &  $-2.66\pm 0.02$ &  $-3.10\pm 0.03$ &  $-3.70\pm 0.07$ &  $-4.33\pm 0.14$ \\ 
\\
$ 1.2 < z \leq 1.5$ & UDS &  $-2.06\pm 0.01$ &  $-2.19\pm 0.01$ &  $-2.38\pm 0.02$ &  $-2.66\pm 0.02$ &  $-3.08\pm 0.03$ &  $-3.84\pm 0.08$ &  $-4.68\pm 0.22$ \\ 
 & Cosmos &  $-2.35\pm 0.03$ &  $-2.48\pm 0.03$ &  $-2.68\pm 0.04$ &  $-2.89\pm 0.05$ &  $-3.29\pm 0.08$ &  $-3.66\pm 0.12$ &  $-4.33\pm 0.25$ \\ 
[+1ex]
 & Total &  $-2.12\pm 0.01$ &  $-2.25\pm 0.01$ &  $-2.44\pm 0.01$ &  $-2.71\pm 0.02$ &  $-3.13\pm 0.03$ &  $-3.79\pm 0.07$ &  $-4.57\pm 0.16$ \\ 
\\
\enddata
\end{deluxetable*}

\section{Cumulative Velocity Dispersion Function}

We also consider the evolution of velocity dispersions of galaxies at constant cumulative number density.  Assuming that the rank order of galaxies remains the same, a number density selection selects the same population of galaxies as a function of redshift (e.g., {Wake} {et~al.} 2008; {van Dokkum} {et~al.} 2010).  We show the cumulative VDF in the bottom row of Figure \ref{fig:VDFcomp} and select galaxies at three number densities: $\log\,N[\unit{Mpc^{-3}}]=-3.5,-4.5,\&-5.5$ (horizontal gray lines).  The crossing point between each number density threshold and the cumulative VDF evolves differently with redshift.

\begin{figure}[!h]
  \centering
%  \epsscale{0.5}
  \plotone{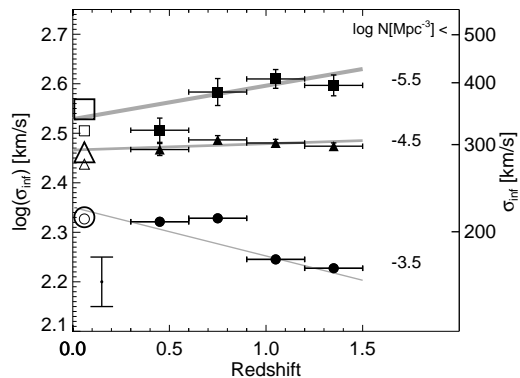}
  \caption{Velocity dispersion at the three number density thresholds ($\log\,N[\unit{Mpc^{-3}}]=-3.5,-4.5,\&-5.5$) shown in the bottom row of Figure \ref{fig:VDFcomp}.  Open symbols represent the unscattered (small symbols) and scattered (large symbols) {Sheth} {et~al.} (2003) total VDF.  Filled symbols represent $\sigma_{cross}$ for the higher redshift VDFs.  Vertical error bars include the $0.06\unit{dex}$ error in inferred velocity dispersion divided by the square root of the number of galaxies interpolated between two nearest bins in $\log\sigma$ from Figure \ref{fig:VDFcomp}.  We include an additional error of $0.05\unit{dex}$ (error bar in lower left corner), added in quadrature to all points, to encompass additional sources of error.}
  \label{fig:ndens_sigmaz}
\end{figure}

In Figure \ref{fig:ndens_sigmaz}, we show the evolution of the velocity dispersion at fixed cumulative number density ($\sigma_{cross}$)
 with redshift.  The rare high-dispersion
galaxies ($\log\,N[\unit{Mpc^{-3}}]<-5.5$) exhibit a weak trend with
even higher dispersions at high redshift and a slope of
$0.066\pm0.059$.  While it is difficult to decrease a galaxy's velocity
dispersion with time, it has been suggested that this can occur via extensive
minor merging, which can efficiently increase size while slowly
increasing mass (e.g., {Bezanson} {et~al.} 2009; {Hopkins} {et~al.} 2009; {van Dokkum} {et~al.} 2010).  An abundance of evidence has demonstrated that at a
given mass, galaxies have smaller sizes at high redshift than in the
local universe (e.g., {Daddi} {et~al.} 2005; {Trujillo} {et~al.} 2006; {van Dokkum} {et~al.} 2008; {Cimatti} {et~al.} 2008; {van der Wel} {et~al.} 2008; {Franx} {et~al.} 2008; {Damjanov} {et~al.} 2009).  We emphasize that these
are the most extreme galaxies in the universe - with inferred velocity
dispersions of $\gtrsim400\,\rm{km/s}$ at $z\sim0$.  The fact that
this trend is weak suggests that the structure correction to the
inferred dispersions may mitigate the evolution in density within
$r_e$.  At the higher threshold of $\log\,N[\unit{Mpc^{-3}}]<-4.5$
galaxies appear to have already attained the central velocity
dispersions that they will have at $z=0$, with a slope of
$0.014\pm0.054$.  If these galaxies maintain constant central
dispersion while continuing to evolve from $z\sim1.5$ to the present
it could be evidence that they are undergoing inside-out growth in
which the central dynamics of the stars are set by some early
processes and the overall structure of the galaxy builds up around the
core or that they undergo size growth at the same rate as mass growth.
Finally, for the galaxies above the highest density threshold of
$\log\,N[\unit{Mpc^{-3}}]<-3.5$ velocity dispersion increases with
time.  In this case the negative slope of $-0.098\pm0.053$ is significant.  This may suggest the importance of gas accretion and
star formation, which could increase the stellar masses of the
galaxies.  

We note that the dispersion function appears to evolve slower than the
stellar mass function:
stellar mass increases by a factor of $\sim2$ for the most extreme
galaxies ($\log\,M_{\star}>11$) from $z\sim2-0$, with even stronger
evolution at lower masses ({Brammer} {et~al.} 2011).

\section{Discussion}

We stress that all of these results are preliminary and rest on a multitude of assumptions.  First of all, our samples of galaxies with size measurements may suffer from incompleteness, especially at the low mass limit.  We have tried to place thresholds above which our sample should be complete, however these limits may be underestimated.  

The majority of galaxy sizes are measured from ground-based imaging, which could be uncertain, especially for compact galaxies at high redshift.  In Figure \ref{fig:groundsizetest} we have shown that the WIRDS/ACS sizes agree quite well with those measured from WFC3, exhibiting a scatter of only $0.1\unit{dex}$. However, at the high dispersion end of the VDF, the accuracy of individual size measurements becomes more important than overall trends.  Future studies would benefit from careful fitting of profiles exclusively based on space-based imaging.  This will also improve the accuracy of our S\'ersic indices.

The most tentative assumption remains that the relation between
inferred and measured velocity dispersions does not evolve.  There are
many factors that contribute to this relation.  Aperture corrections
depend on galaxy profiles and therefore any systematic evolution in
the light distribution of galaxies would affect the measured velocity
dispersions.  However, this is a relatively weak effect as the
aperture corrections are small.  More important may be the dependence
of $K_v$ on S\'ersic index, which could evolve if galaxy structure
evolves in complex ways.
Furthermore, our calibration of $K_{\star}(n)$
depends strongly on the average ratio of stellar to total dynamical
mass, which may evolve with time and depend on the stellar mass
of the galaxy.  {Taylor} {et~al.} (2010) found that $M_{\star}/M_{dyn,n}$
still exhibits trends with stellar mass at fixed $n$, suggesting that
even at $z=0$ the models for $K_v(n)$ do not completely describe the
non-homology of galaxies.  Any of these effects could preferentially
influence certain types of galaxies, introducing systematic trends in
the shape of the $\sigma_{0}$ vs. $\sigma_{inf}$ relation.  The
scatter in the relation may also evolve with time; this could explain
the apparent increase in number of galaxies with high inferred
dispersions at $z\gtrsim0.6$.  Finally, we assume a constant
{Chabrier} (2003) IMF, which could also differ with redshift or halo
mass (e.g., {van Dokkum} \& {Conroy} 2010).

Measuring the entire VDF using dynamical measurements would require an extremely large spectroscopic survey.  A more economical approach is to better calibrate the evolution of the inferred VDF.  With a large sample of uniformly selected galaxies spanning a large range of velocity dispersions and lower measurement errors, one could characterize $\sigma_{inf}$ as a function of stellar mass, S\'ersic fits and redshift.

Finally, we note that {Chae} (2010) measured the VDF in a
similar redshift range ($z\sim0.3-1.0$) from 
30 lensing galaxies and found quite different results.  Chae
finds that at the low dispersion end ($\log\sigma\lesssim2.3$) the VDF
of galaxies agrees quite well with the local VDF, and that it steepens
at higher redshift.  At the high dispersion end, the {Chae} (2010) results are
based on a lack of arcs with large separations.  It is possible that
this sample of lensing galaxies preferentially misses such lenses.
Alternatively, it is possible that the
intrinsic VDF does fall below the local VDF at high dispersions, but
the scatter increases rapidly with redshift, producing the opposite
evolution in our study.  Further calibration of the inferred velocity
dispersions at higher redshift is necessary to probe whether the two
results are inconsistent.

Although measurements of the VDF at higher redshift may never be as
precise as those measured from the SDSS,  study of the
evolution of the inferred velocity dispersion function using more
accurate and redshift-dependent dynamical calibrations will be a
powerful tool for studying the growth and evolution of galaxies and the
halos that they occupy.

\begin{acknowledgements}
We thank the referee for his/her feedback and acknowledge support from NSF grant AST-0807974 and HST grant GO-12177.01A.
\end{acknowledgements}

%% \bibliography


\begin{references}

\reference{} {Abazajian}, K.~N., {et~al.} 2009, \apjs, 182, 543

\reference{} {Bernardi}, M., {et~al.} 2003, \aj, 125, 1866

\reference{} {Bertin}, G., {Ciotti}, L., \& {Del Principe}, M. 2002, \aap, 386, 149

\reference{} {Bezanson}, R., {van Dokkum}, P.~G., {Tal}, T., {Marchesini}, D., {Kriek}, M.,  {Franx}, M., \& {Coppi}, P. 2009, \apj, 697, 1290

\reference{} {Blanton}, M.~R., {et~al.} 2005, \aj, 129, 2562

\reference{} {Brammer}, G.~B., {van Dokkum}, P.~G., \& {Coppi}, P. 2008, \apj, 686, 1503

\reference{} {Brammer}, G.~B., {et~al.} 2011, ArXiv e-prints

\reference{} {Brinchmann}, J., {Charlot}, S., {White}, S.~D.~M., {Tremonti}, C.,  {Kauffmann}, G., {Heckman}, T., \& {Brinkmann}, J. 2004, \mnras, 351, 1151

\reference{} {Bruzual}, G., \& {Charlot}, S. 2003, \mnras, 344, 1000

\reference{} {Cappellari}, M., {et~al.} 2006, \mnras, 366, 1126

\reference{} ---. 2009, \apjl, 704, L34

\reference{} {Chabrier}, G. 2003, \pasp, 115, 763

\reference{} {Chae}, K. 2010, \mnras, 402, 2031

\reference{} {Choi}, Y., {Park}, C., \& {Vogeley}, M.~S. 2007, \apj, 658, 884

\reference{} {Cimatti}, A., {et~al.} 2008, \aap, 482, 21

\reference{} {Daddi}, E., {et~al.} 2005, \apj, 626, 680

\reference{} {Damjanov}, I., {et~al.} 2009, \apj, 695, 101

\reference{} {Djorgovski}, S., \& {Davis}, M. 1987, \apj, 313, 59

\reference{} {Franx}, M., {van Dokkum}, P.~G., {Schreiber}, N.~M.~F., {Wuyts}, S.,  {Labb{\'e}}, I., \& {Toft}, S. 2008, \apj, 688, 770

\reference{} {Hopkins}, P.~F., {Bundy}, K., {Murray}, N., {Quataert}, E., {Lauer}, T.~R., \&  {Ma}, C. 2009, \mnras, 398, 898

\reference{} {Kriek}, M., {van Dokkum}, P.~G., {Labb{\'e}}, I., {Franx}, M., {Illingworth},  G.~D., {Marchesini}, D., \& {Quadri}, R.~F. 2009, \apj, 700, 221

\reference{} {Lawrence}, A., {et~al.} 2007, \mnras, 379, 1599

\reference{} {Loeb}, A., \& {Peebles}, P.~J.~E. 2003, \apj, 589, 29

\reference{} {Lonsdale}, C.~J., {et~al.} 2003, \pasp, 115, 897

\reference{} {Magorrian}, J., {et~al.} 1998, \aj, 115, 2285

\reference{} {Mitchell}, J.~L., {Keeton}, C.~R., {Frieman}, J.~A., \& {Sheth}, R.~K. 2005,  \apj, 622, 81

\reference{} {Naab}, T., {Johansson}, P.~H., \& {Ostriker}, J.~P. 2009, \apjl, 699, L178

\reference{} {Newman}, A.~B., {Ellis}, R.~S., {Treu}, T., \& {Bundy}, K. 2010, \apjl, 717,  L103

\reference{} {Onodera}, M., {et~al.} 2010, \apjl, 715, L6

\reference{} {Peng}, C.~Y., {Ho}, L.~C., {Impey}, C.~D., \& {Rix}, H.-W. 2002, \aj, 124, 266

\reference{} {Schechter}, P. 1976, \apj, 203, 297

\reference{} {Scoville}, N., {et~al.} 2007, \apjs, 172, 1

\reference{} {Sekiguchi}, K., \& {SXDS}. 2004, in Bulletin of the American Astronomical  Society, Vol.~36, American Astronomical Society Meeting Abstracts, 1478--+

\reference{} {S\'ersic}, J.~L. 1968, {Atlas de galaxias australes} (Cordoba, Argentina:  Observatorio Astronomico, 1968)

\reference{} {Sheth}, R.~K., {et~al.} 2003, \apj, 594, 225

\reference{} {Taylor}, E.~N., {Franx}, M., {Brinchmann}, J., {van der Wel}, A., \& {van  Dokkum}, P.~G. 2010, \apj, 722, 1

\reference{} {Trujillo}, I., {Ferreras}, I., \& {de la Rosa}, I.~G. 2011, ArXiv e-prints

\reference{} {Trujillo}, I., {et~al.} 2006, \apj, 650, 18

\reference{} {van de Sande}, J., {et~al.} 2011, ArXiv e-prints

\reference{} {van der Wel}, A., {Holden}, B.~P., {Zirm}, A.~W., {Franx}, M., {Rettura}, A.,  {Illingworth}, G.~D., \& {Ford}, H.~C. 2008, \apj, 688, 48

\reference{} {van Dokkum}, P.~G., \& {Conroy}, C. 2010, \nat, 468, 940

\reference{} {van Dokkum}, P.~G., {Kriek}, M., \& {Franx}, M. 2009, \nat, 460, 717

\reference{} {van Dokkum}, P.~G., {et~al.} 2008, \apjl, 677, L5

\reference{} ---. 2010, \apj, 709, 1018

\reference{} {Wake}, D.~A., {et~al.} 2008, \mnras, 387, 1045

\reference{} {Warren}, S.~J., {et~al.} 2007, \mnras, 375, 213

\reference{} {Whitaker}, K.~E., {et~al.} 2011, ArXiv e-prints

\reference{} {Williams}, R.~J., {Quadri}, R.~F., {Franx}, M., {van Dokkum}, P., \&  {Labb{\'e}}, I. 2009, \apj, 691, 1879

\reference{} {Williams}, R.~J., {Quadri}, R.~F., {Franx}, M., {van Dokkum}, P., {Toft}, S.,  {Kriek}, M., \& {Labb{\'e}}, I. 2010, \apj, 713, 738

\end{references}
\end{document}